\documentclass[letterpaper]{article}

\usepackage{natbib}
\usepackage{alifeconf} % Order is important

\usepackage{url}
\usepackage{xcolor}
\definecolor{tealcite}{rgb}{0.0,0.45,0.5}
\usepackage[colorlinks,citecolor=tealcite,urlcolor=magenta]{hyperref}
\usepackage{xurl}
\usepackage{amsmath}
\usepackage{amssymb}
\usepackage{amsfonts}
\usepackage{float}
\usepackage[normalem]{ulem}
\usepackage{titlesec}
\usepackage{enumitem}
\setlist{nosep}
\titlespacing*{\section}{0pt}{1.5ex}{0.8ex}
\titlespacing*{\subsection}{0pt}{1.0ex}{0.5ex}
\titlespacing*{\paragraph}{0pt}{0pt}{1em}

\title{Agnosiophobia in a virtual agent: behavioral and dynamical architecture in Lenia}

\author{
    Jesse Cool$^{1}$,
    Benedikt Hartl$^{2}$,
    Michael Levin$^{2,3,*}$,
    Samantha Petti$^{1,*}$\\
    \mbox{}\\
    $^1$Department of Mathematics, Tufts University \\
    $^2$Allen Discovery Center, Tufts University\\
    $^3$Wyss Institute for Biologically Inspired Engineering, Harvard University\\
    \mbox{}\\
    $^*$Corresponding authors: michael.levin@tufts.edu, spetti@tufts.edu
}

\begin{document}

\maketitle

\begin{abstract}
All embodied agents are fundamentally patterns in physiological or other excitable media, blurring the distinction between objects and processes. Emergent patterns with complex behaviors, such as Gliders in the Game of Life and virtual patterns in Lenia, are powerful model systems in which to understand the properties and origins of behavioral traits in novel agents. To evaluate the behavior of patterns in Lenia, we introduce regions into their environment from which no sensory information is available -- in effect, making creatures blind to parts of their surroundings. Complementing the conventional concept of infotaxis, we find that creatures tend to avoid these regions, a behavior we term \textit{agnosiophobia}. To explain this behavior, we map each test creature's sensitivity to targeted occlusions and interpret the results in the language of dynamical systems. We observe Lenia creatures taking advantage of their freedom to change heading in order to achieve what appears to be a more fundamental goal: the preservation of their morphology. This work illustrates the beginning of an important roadmap to understand how emergent agents’ behavioral propensities interact with the informational, not only tangible, topography of their world.
\end{abstract}

\begin{center}
\href{https://github.com/jessescool/lenia-umwelt}{\texttt{github.com/jessescool/lenia-umwelt}}
\end{center}

\section{Introduction}

Conventional biology, as well as cognitive science, tends to make a clear distinction between thoughts and thinkers, data and machines that process it, and real embodied agents vs. dynamic patterns in excitable media. However, this distinction is in the eye of the beholder, depending heavily on the spatial and temporal scales of an observer, as well as their sensory capabilities. Even we humans, the prototypical example of a real embodied being, are, in an important sense, temporary metabolic patterns \citep{Levin2019, Levin2022, Dennett2020}. 

% Likewise, observing organisms through the lens of gene expression, physical force, or physiological state reveals entirely different boundaries and scales of coherent subsystems. These subsystems navigate those spaces, perform niche construction to facilitate their own existence, hack each other with stimuli, perform computations, and pursue homeodynamic setpoints with diverse degrees of competency \citep{Levin2019, Levin2022, Dennett2020}.

\begin{figure}[t]
    \centering
    \includegraphics[width=\columnwidth]{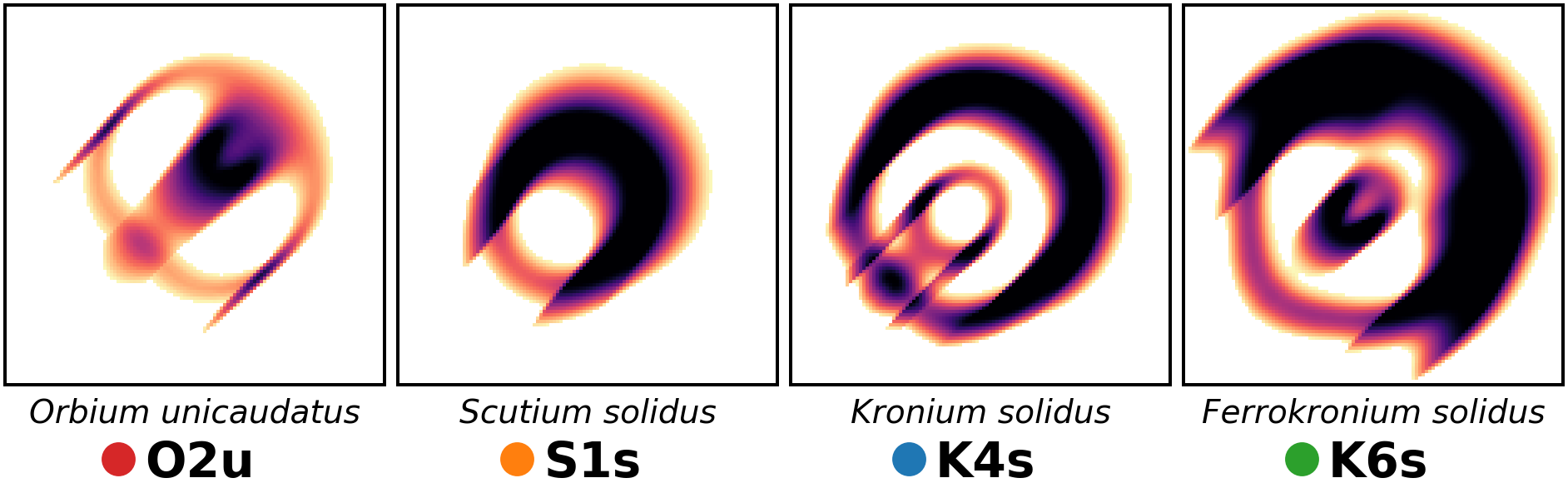}
    \caption{\textbf{Four stable Lenia creatures} \citep{Chan2019}, each a self-maintaining motile pattern that emerges from a particular ruleset and initial condition. Intensity represents cell activation value in $[0,1]$; active cells range from light orange (low) to black (high). Species names and abbreviations follow \citet{Chan2019}. Colored dots identify each creature throughout all figures in this paper.}
    \label{fig:creatures}
\end{figure}

Thus, it has been argued that the distinction between patterns and objects is a continuum, not a fundamental category \citep{Fields2025}. This is compatible with the general tenets of Artificial Life, expanding the range of systems that may be studied using the tools of the behavioral, cybernetic, and informational sciences. This also has many implications for the origin of life \citep{Küppers1990, Mirmomeni2014}, the biomedicine of physiological patterns (which can be causal agents in health and disease, alongside physical patterns such as cellular structures), behavioral science (via the dynamics of information content in cognitive systems), exobiology (the need to recognize highly unconventional forms of life and mind), and robotics/AI \citep{Braitenberg2004, Kriegman2020}. 

A key question for any agent is where it fits on the spectrum of agency \citep{Dennett1989, Barandiaran2009, Rosenblueth1943, Fields2022}: what is important is the degree of competency it has in forming and pursuing goals -- its creative capacity for dealing with problems and opportunities -- regardless of its implementation and its degree of conventional solidity in the 3D world. This is especially critical for novel beings, whether created by synthetic bioengineering \citep{Clawson2023} or artificial life efforts \citep{Braitenberg2004, Kriegman2020}. In this paper, we probe for protocognition in \textit{Lenia} \citep{Chan2019}, a continuous cellular automaton, by subjecting its self-maintaining motile patterns to informational occlusions.

Emergent dynamics in minimal computational media are powerful model systems with which to expand our frameworks for understanding alien minds \citep{Sloman1984, Shanahan2010, Baluška2016}. A familiar example of such patterns is the Glider in Conway’s Game of Life \citep{Conway1970}. In a classic study, \citet{Beer2014} analyzed its Umwelt -- the world as experienced through its own sensory apparatus -- by exhaustively classifying all $2^{24}$ perturbations according to the structural transitions they induced, including changes in form, chirality, and direction of motion. Previously, we showed surprising latent competencies in very simple systems -- sorting algorithms -- that found novel ways to complete their task when tangible barriers were placed in their way \citep{Zhang2025}. Inspired by but not analogous to these approaches, we investigate a more subtle response, indicative of a more sophisticated level of protocognition, in Chan's Lenia \citep{Chan2019}. 

\textit{Lenia} is a continuous cellular automaton whose spatially-uniform update rule produces a slew of emergent dynamics \citep{Chan2019}. From a subset of rulesets and initial conditions arise self-maintaining motile structures that propagate like solitons, localized patterns that travel while preserving their approximate form~(Figure~\ref{fig:creatures}). Though the rules of Lenia are symmetric under translation and rotation, these patterns are often not: each occupies a particular position and travels in a particular direction, breaking the symmetries of the underlying dynamics.

Since its introduction, Lenia has been generalized to multiple channels and higher dimensions \citep{Chan2020}, implemented on alternative substrates \citep{Davis2024, Kawaguchi2021}, and used as a testbed for open-ended evolution and the automated discovery of virtual patterns \citep{Reinke2020, Faldor2024, Kumar2025, Hamon2025}. Others have explored the structure of Lenia's parameter space \citep{Hudcová2026} and trained RL agents to guide Lenia creatures toward target states \citep{Cvjetko2025}. Less attention has been paid to the internal structure of the creatures themselves.

We subject four Lenia creatures to informational occlusions, assessing the competency of these dynamic patterns as they navigate novel environments, motivated by the question of where they fall on the spectrum of agency. First, we review the structure of Lenia and motivate the form of our perturbation. We then document a behavior we term ``agnosiophobia'' -- fear of the unknown -- across creatures as they traverse environments with varied regions of occlusion. These creatures avoid occluded regions despite having no explicit mechanism designed for the task. To probe this mechanism, we map each creature's vulnerability and resilience to targeted occlusions, illustrating how perturbations can turn morphological resilience into navigational capacity. Finally, we interpret Lenia as a dynamical system, using this framing to explain the observed behavior and connect our findings to broader questions of goal-directedness.

\section{System and Methods}

In Lenia, cells on a toroidal grid take values in $[0,1]$ \citep{Chan2019, Chan2020}. The system evolves via a global, memoryless iterative map $F: A_t \mapsto A_{t+1}$ defined by local update rules applied uniformly and in parallel across all cells.

Each cell in Lenia computes its next state by convolving its neighborhood with a normalized, radially symmetric kernel $K$ whose entries sum to 1; the potential of each grid cell $a$ is given by $U_t(a) = (K * A_t)(a)$, where $*$ represents convolution. $U_t$ is then passed through a growth function:

\begin{equation}
A_{t+1} = \text{clip}_{[0,1]}\bigl(A_t + \Delta t \cdot G(U_t)\bigr)
\label{eq:update}
\end{equation}

$G$ is a fixed function, specific to each ruleset, that maps neighborhood potential to cell growth or decay. $\Delta t$ represents step size, controlling the granularity of the update.

\subsection{Update Rule with Occlusion}

The kernel encodes the cell's sensory field -- what it can `see' and how much it `cares' \citep{Chan2019}. This extends naturally to what the creature can `see': sensory information propagates quickly through cells as a simulation is stepped through time. 

Our intervention targets $U_t$ by modifying what information is observed by the kernel. We occlude regions of the grid, excluding them from the convolution operation. Occluded regions do not take values in $[0,1]$ -- such values would be \textit{observed} and subsequently factored into the cell's next state. We represent them as a binary mask $B$ ($B_{ij} = 1$ occluded, $0$ visible) matching the dimensions of $A$.

\begin{equation}
U_t = \frac{K * \bigl(A_t \cdot (\mathbf{1} - B)\bigr)}{K * (\mathbf{1} - B)}
\label{eq:renorm}
\end{equation}

The numerator convolves over all cells but first sets masked cells to $0$. Formally, multiplying the state $A_t$ by $(\mathbf{1}-B)$ imposes a pixel-wise Heaviside filter defined by the field $B$, where visible input passes unchanged while occluded regions are fully suppressed. The denominator renormalizes by the fraction of kernel mass that lands on visible territory, preventing occluded cells from contributing spurious zeroes to $U_t$.\footnote{Some might recognize this operation as analogous to rescaling after dropout \citep{Srivastava2014}.} When a cell's kernel lies entirely within visible territory, Equation~\ref{eq:renorm} reduces to the standard update rule. This is a minimal, behaviorally-informed intervention: \textit{let the remaining information sway us more, proportional to how much is obscured}.

\subsection{Measuring recovery and its quality}

\begin{figure}[t]
    \centering
    \includegraphics[width=\columnwidth]{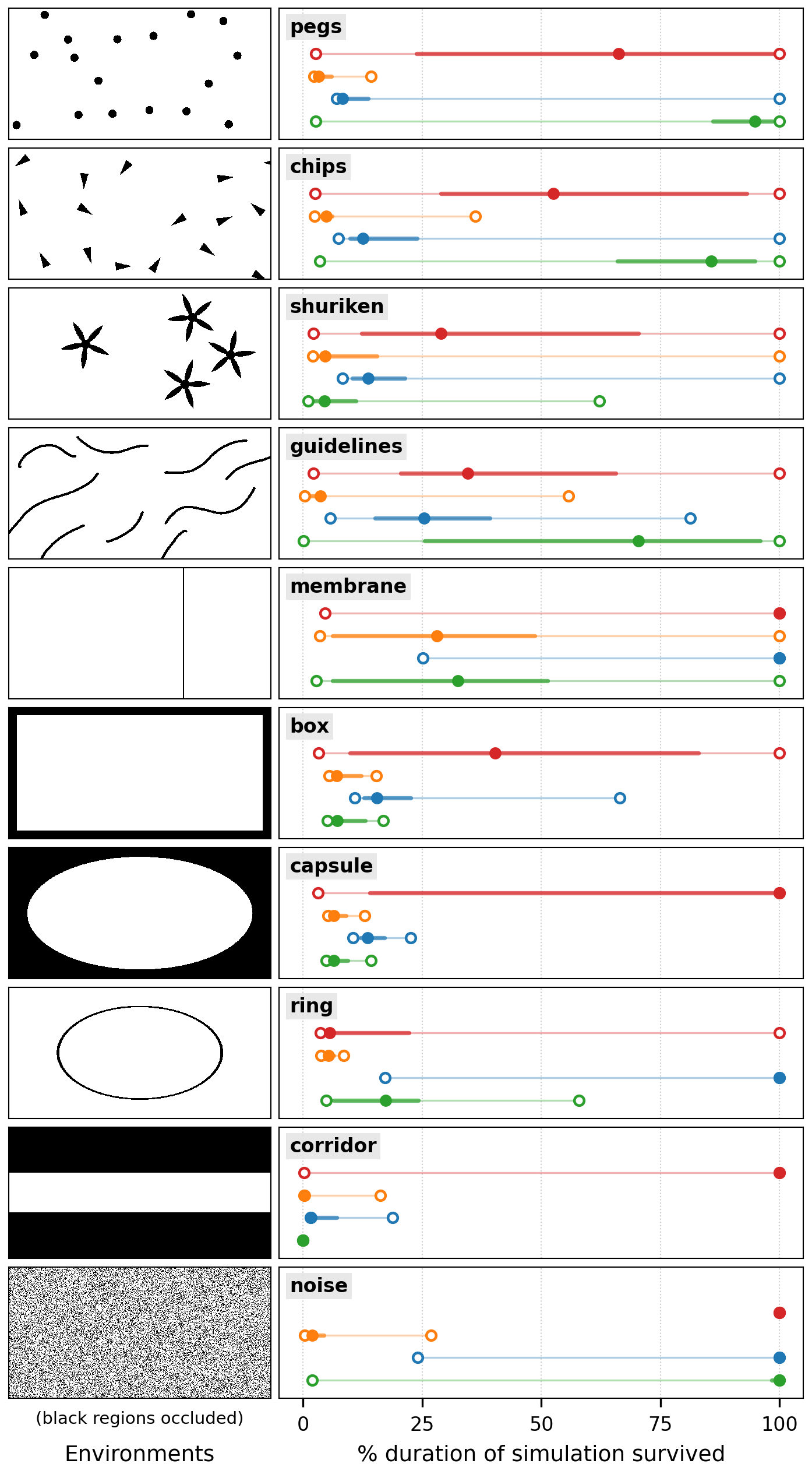}
    \caption{\textbf{Survival competence of creatures by environment.} Each creature starts at the center of each environment facing each of 360 orientations. We run creatures for $2000$ steps, enough to observe variation in average survival rates by creature. Black regions are occluded. Bars show the distribution of percentage of simulation time survived, with median, inter-quartile range, and extremes. Colors correspond to the creatures in Figure~\ref{fig:creatures}.}
    \label{fig:env_competence}
\end{figure}

\begin{figure*}[!tp]
    \centering
    \includegraphics[width=\textwidth]{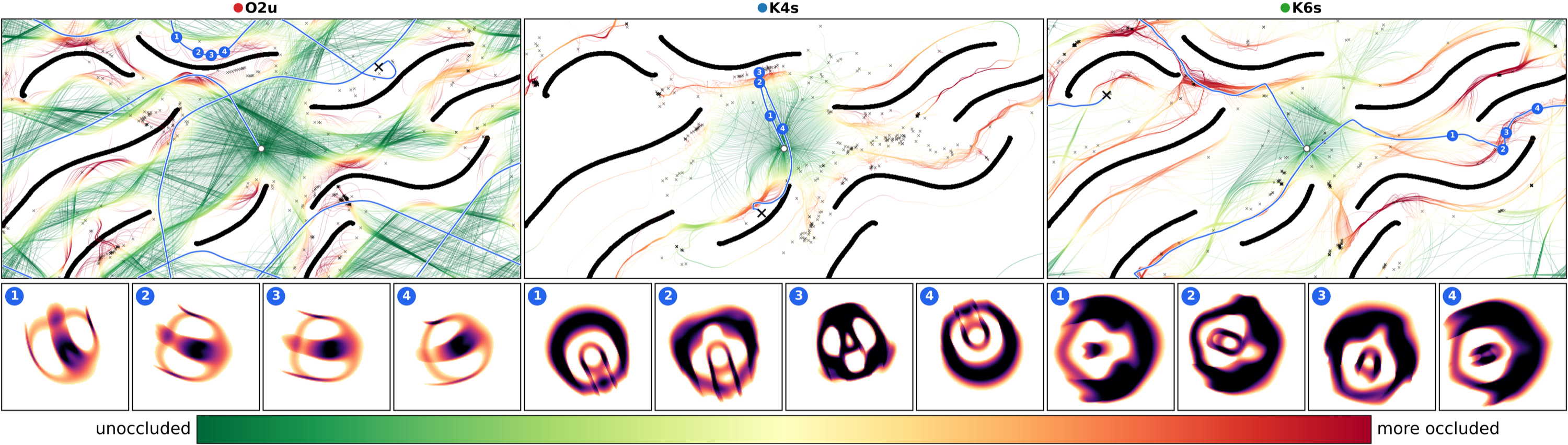}
    \caption{\textbf{Agnosiophobia in three Lenia creatures.} O2u, K4s, and K6s (Figure~\ref{fig:creatures}) navigating the \textit{guidelines} environment (Figure~\ref{fig:env_competence}). Each line represents the trajectory of one of 360 orientations. Color represents the fraction of kernel area occluded across all nonzero pixels. We highlight one path per creature, where $\times$ marks the last frame at which the creature was within its neighborhood, marking approximate death. S1s is omitted as it rarely survives long enough to produce meaningful trajectories.}
    \label{fig:trajectory_overlay}
\end{figure*}

Lenia creatures occupy a continuous region of the state space of a simulation, rather than a single configuration. At finite resolutions, they are not perfectly translationally and rotationally invariant copies of themselves: their pixel-level representations fluctuate slightly at each step, and differently so depending on the creature's angle relative to the axes of grid symmetry. Where a creature begins and ends is not established \textit{a priori}. Hence, there is no `perfect circle' dividing those configurations considered to be a creature from the rest. 

\textbf{A Wasserstein distance for comparing Lenia creatures.} To determine whether two grid states contain instances of the `same' creature, we construct a distance metric, invariant to translation and rotation, meant to capture morphology. We compare grid states by distributions of active pixels, what we call their \textit{profile}: we flatten each grid, sort nonzero values in descending order, and trim or zero-pad to a common length $m$. The distance between two states becomes the Wasserstein-1 distance between their profiles:

\begin{equation}
\begin{aligned}
d(a,b) &= W_1\!\bigl(\mathrm{profile}(a),\;\mathrm{profile}(b)\bigr) \\
       &= \frac{1}{m}\sum_{i=1}^{m}\bigl|a_i - b_i\bigr|
\end{aligned}
\label{eq:distance}
\end{equation}

Though not injective, this metric reliably separates the classes we are concerned with: the creature, the blank grid, explosion\footnote{`Explosion' is the resulting class of grid states achieved after unbounded growth due to perturbation. Explosions scatter regions of high mass across the grid, far from any creature's stable morphological pattern \citep{Chan2019}.}, and metamorphosis.

\textbf{Neighborhood.} While the Wasserstein distance (Equation~\ref{eq:distance}) between two instances of the same creature is small, it is usually non-zero. We define a creature's \textit{neighborhood} to capture per-species variation in $d$, an empirical measure of its natural morphological variability.

For each creature, we construct a dataset $C$ of 5400 snapshots, drawn from simulations initialized at 90 different orientations, each run for 600 time steps. This samples the creature's canonical form across a range of phases and orientations. We compute a barycenter $\bar{c}$ as the element-wise median of their profiles:

$$ \bar{c}_i = \mathrm{median}\bigl(\{\text{profile}(c_i) \mid c \in C\}\bigr) $$

This is the Wasserstein-1 barycenter, an exemplar of the activation profile of a creature. We define $d_{\max} = \max(\{d(\bar{c}, c) \mid c \in C\})$, the furthest any observed canonical pattern strays from $\bar{c}$. We define the \textit{neighborhood} of a creature $\mathcal{N}(\bar{c},\, d_{\max})$ to be the set of morphologies within $d_{\max}$ of $\bar{c}$, the reference against which recovery is measured.

\textbf{Quantifying recovery.} For each frame $A_t$ of a post-perturbation trajectory, we compute $d(A_t, \bar{c})$, yielding a time-series of Wasserstein-1 distances. We label a creature morphologically \textbf{recovered} at time $t$ if the average distance over the past $k$ (default 5) consecutive frames remains below $d_{\max}$, and say that it has `returned to its neighborhood.' A creature is \textbf{dead} if its total mass drops below $0.01$ at any frame. Otherwise, we say a creature has \textbf{not recovered}, representing explosion and metamorphosis. These three outcomes yield the results presented in Figures~\ref{fig:env_competence}, \ref{fig:trajectory_overlay}, \ref{fig:sweeps}, and \ref{fig:long_windy_recovery}. We characterize those trajectories that result in recovery:

\begin{itemize}
    \item{\textbf{Frames until recovery} -- after a perturbation, the number of time steps before the next within-neighborhood frame.}
    \item{\textbf{Max distortion} -- the maximum $d$ between a creature's $\bar{c}$ and each frame along the recovery trajectory.}
    \item{\textbf{Heading change} -- the angle by which heading changes between pre-perturbation and post-recovery trajectories, derived from mass-weighted centroid paths.}
\end{itemize}

\noindent We select four stable, non-oscillating, motile creatures from Chan's catalog (Figure~\ref{fig:creatures}) for testing, each defined in a $(K, G, \Delta t)$ ruleset. These creatures are upscaled to higher resolution than Chan's originals, to allow finer-grained perturbations at a resolution closer to the continuous limit.

\section{Lenia creatures avoid regions of occlusion}
\label{sec:environments}

To determine how Lenia creatures' behavior relates to the sensory information available in their environment, we placed creatures in environments containing occluded regions -- spatial masks $B$ (Equation~\ref{eq:renorm}) from which no information reaches the creature. These regions vary in size, sparsity, convexity/concavity, and slope. We ran each of our four test creatures (Figure~\ref{fig:creatures}) in 10 environments (Figure~\ref{fig:env_competence}), starting in the center facing each of 360 directions.

When a creature's sensory field overlaps with an occluded region, the resulting perturbation to kernel potential disrupts its dynamics. Depending on the extent and location of occlusion, this can push the creature toward death, metamorphosis, or explosion. Despite this, and despite having no mechanism designed for the task, we observed a pattern we call \textit{agnosiophobia} – reorientation away from these regions – among three of our creatures: O2u, K4s, and K6s. S1s did not exhibit this capacity and died upon encountering occlusions. Figure~\ref{fig:trajectory_overlay} illustrates this behavior for three creatures in the \textit{guidelines} environment. Survival rates varied substantially by creature.

\textbf{O2u.} Of our four test creatures, O2u (the Orbium) demonstrates the most robust agnosiophobia (Figure~\ref{fig:trajectory_overlay}). Its trajectory remains straight until the kernel fields of its active pixels begin to overlap with occluded regions; sensing one on its front right flank, it begins to rotate left, more strongly as it draws nearer. O2u redirects most effectively when it encounters occlusions asymmetrically (not `head-on,' but from an angle; observe trajectories in Figure~\ref{fig:trajectory_overlay}). It performs well in most environments (Figure~\ref{fig:env_competence}).

\textbf{K4s.} On average, K4s exhibits a lesser capacity to survive across our environments than O2u and K6s (Figure~\ref{fig:env_competence}). When it persists, it often skirts the edge of an occluded region and, upon reaching its end, launches away in a new direction. In some encounters with occlusions at its immediate front, K4s transforms into a distinct shape, an oscillating structure, and re-emits in the opposite direction (Figure~\ref{fig:trajectory_overlay}).

\textbf{K6s.} K6s does not always redirect away from occlusions; it often travels along their edges rather than turning away immediately, producing the skirting behavior visible in Figure~\ref{fig:trajectory_overlay}. K6s endures lengthy durations of significant occlusion (Figures~\ref{fig:trajectory_overlay} and~\ref{fig:env_competence}).

Figure~\ref{fig:trajectory_overlay} illustrates trajectories in the \textit{guidelines} environment, which we chose for the visual clarity of its trajectories and the range of approach angles it affords. We observed agnosiophobia, however, across all ten environments (Figure~\ref{fig:env_competence}). O2u survived the longest on average, S1s the shortest. These creatures share the same general update framework and style of renormalized kernel and yet produce qualitatively different behavioral responses -- each a consequence of their respective morphologies and the rules governing their dynamics. To further understand the interaction between occluded regions and a creature's sensory and response apparatus, we target specific locations within each creature's morphology and measure the resulting recovery.

\section{Sensitivity to occlusion is spatially structured}

\label{sec:sweeps}

A defining feature of agents is how they demarcate the boundary between self and environment, distinguishing their inner components and processes from those of the `outside world' \citep{Levin2019}. Our test creatures have a defined structural border that moves in their universe. In our environments (Figure~\ref{fig:env_competence}), creatures sense the dearth of information at a distance: a cell detects occlusion as soon as its kernel -- which extends $R$ pixels radially -- overlaps with an occluded region. Because $R$ varies by creature, so does the distance at which course-corrections begin, well before any of the creature's own `body' is directly occluded (Figure~\ref{fig:trajectory_overlay}). 

\begin{figure}[t]
    \centering
    \includegraphics[width=\columnwidth]{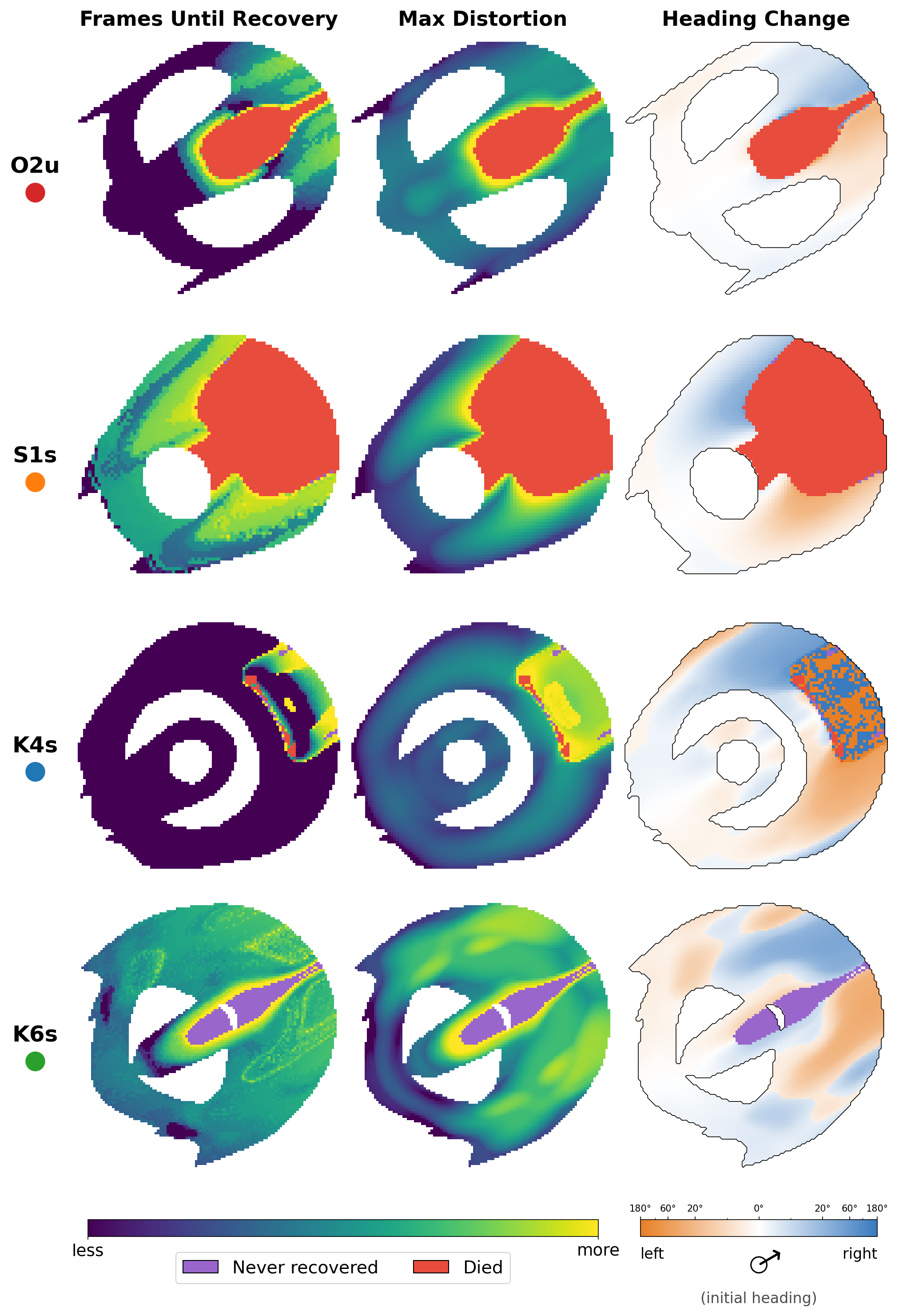}
    \caption{\textbf{Behavioral and recovery response by targeted intervention location.} For each nonzero pixel across four creatures (heading at 2 o'clock), we place a permanent $3\times$3 occluded region and record the resulting dynamics. Left: number of frames until recovery. Center: maximum morphological distortion during recovery. Right: heading change relative to initial heading (blue = left, red = right); note that dark blue and dark orange both represent near-$180 ^\circ$ rotation. Red indicates death; purple indicates explosion or metamorphosis. For frames to recovery and max distortion plots, identical colors across creatures do not indicate identical recoveries; colors are normalized by creature. These heatmaps are consistent across heading directions.}
    \label{fig:sweeps}
\end{figure}

To determine how creatures react to occlusions within their own structure, we place a persistent $3 \times 3$ occluded region at each nonzero pixel of a creature's mass and record the resulting dynamics (Figure~\ref{fig:sweeps}). We observed markedly different sensitivity profiles among our four test creatures.

Across all four creatures, the approximate response is consistent. Perturbations to the right flank cause leftward reorientation, and vice versa. Regions most sensitive to reorientation flank and abut lethal or non-recoverable regions, forming a gradient: perturbations closer to lethal regions have the capacity to provoke larger heading changes via longer, more distorted recovery paths (Figure~\ref{fig:sweeps}). Creatures differ in how this structure is organized.

Perturbations lethal to O2u are concentrated thinly at the center of its leading edge and expand into its core. Preceding them, in the direction of travel, are regions sensitive to reorientation, some that provoke near $180^\circ$ heading change. 

\begin{figure*}[t!]
    \centering
    \includegraphics[width=\textwidth]{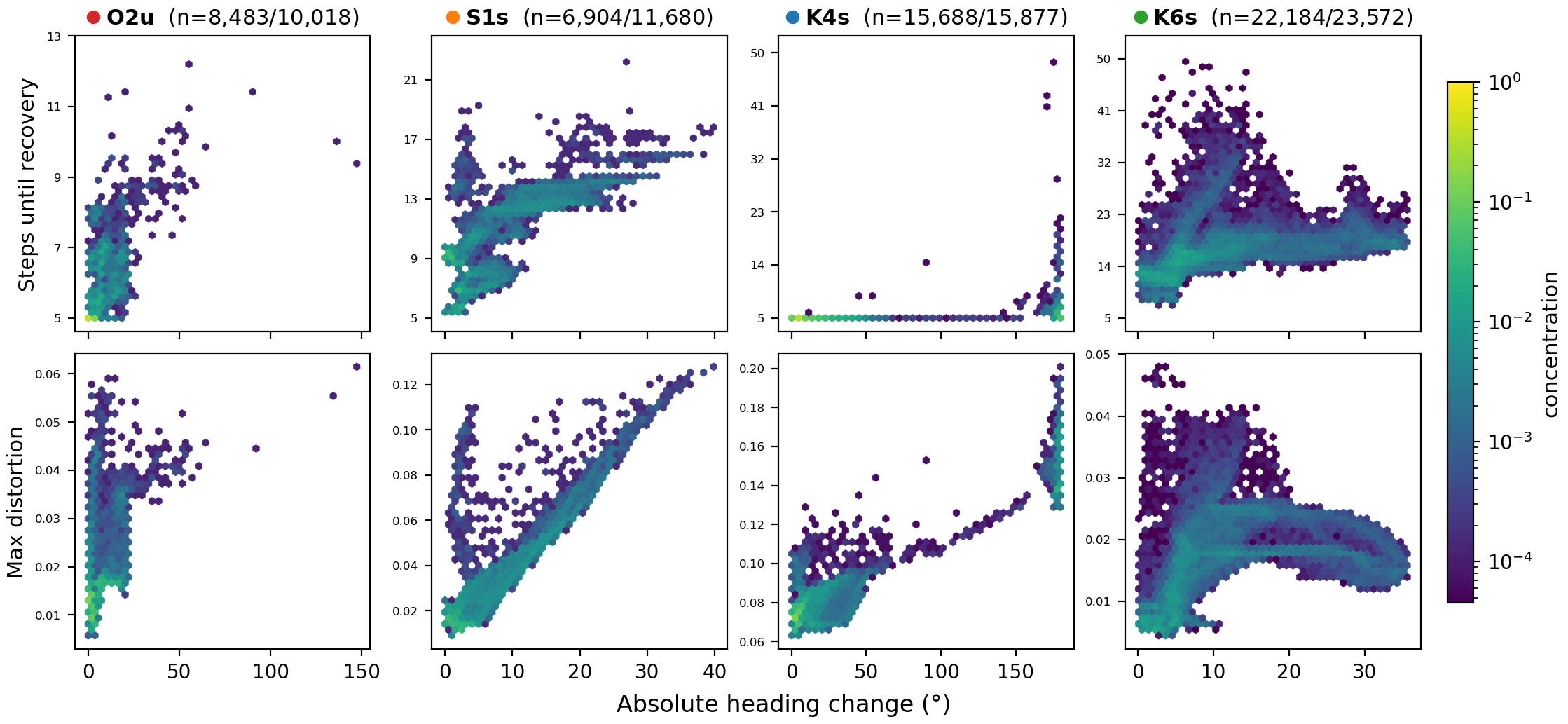}
    \caption{\textbf{Greater heading change requires long and distorted recovery.} Each point represents one recovered trajectory from the sweeps in Figure~\ref{fig:sweeps}, aggregated across 4 orientations. For each creature, $n=a/b$ denotes $a$ recovered perturbations out of $b$ total nonzero pixels across all 4 orientations. Color encodes frequency of observation. Lower-right regions tend to be empty: creatures rarely achieve large heading change quickly and without distortion.}
    \label{fig:long_windy_recovery}
\end{figure*}

S1s, by contrast, is extremely fragile. Interventions are lethal along the majority of its front region, leaving little chance for an environmental occlusion to disturb only the regions that produce reorientation. S1s' poor survival across our environments (Figure~\ref{fig:env_competence}) follows from this fragility. 

The sparse lethal regions along K4s' body exist behind regions that provoke significant reorientation. Its heading change map has a sharp threshold between perturbations that cause soft reorientation and those that invert the creature's direction (Figure~\ref{fig:sweeps}), visible in Figure~\ref{fig:trajectory_overlay}. Nearby perturbation locations can produce drastically different changes in heading. The core and rear of K4s is robust, recovering quickly with small morphological distortion.

None of our targeted occlusions are lethal to K6s. Some, along its axis of symmetry, result in metamorphosis and explosion (Figure~\ref{fig:sweeps}). Its heading changes are comparatively weak, consistent with its skirting behavior in environments (Figure~\ref{fig:trajectory_overlay}), and its recovery behavior is relatively uniform across its body.

In Figure~\ref{fig:long_windy_recovery}, we observe that locations that provoke large heading changes exist exclusively in the higher time-to-recovery, higher-distortion region. \uline{No significant reorientation results from quick, undistorted recoveries.} Distorted and often long recovery appears necessary for reorientation, but not sufficient: many long, distorted recovery paths produce little change in heading. This suggests that navigational capacity and fragility go hand-in-hand -- perturbations capable of causing large heading changes push the creature through extended, morphologically-distorted states, close to the boundary of survival.

\section{Lenia as a dynamical system}

\begin{figure}[t]
    \centering
    \includegraphics[width=\columnwidth]{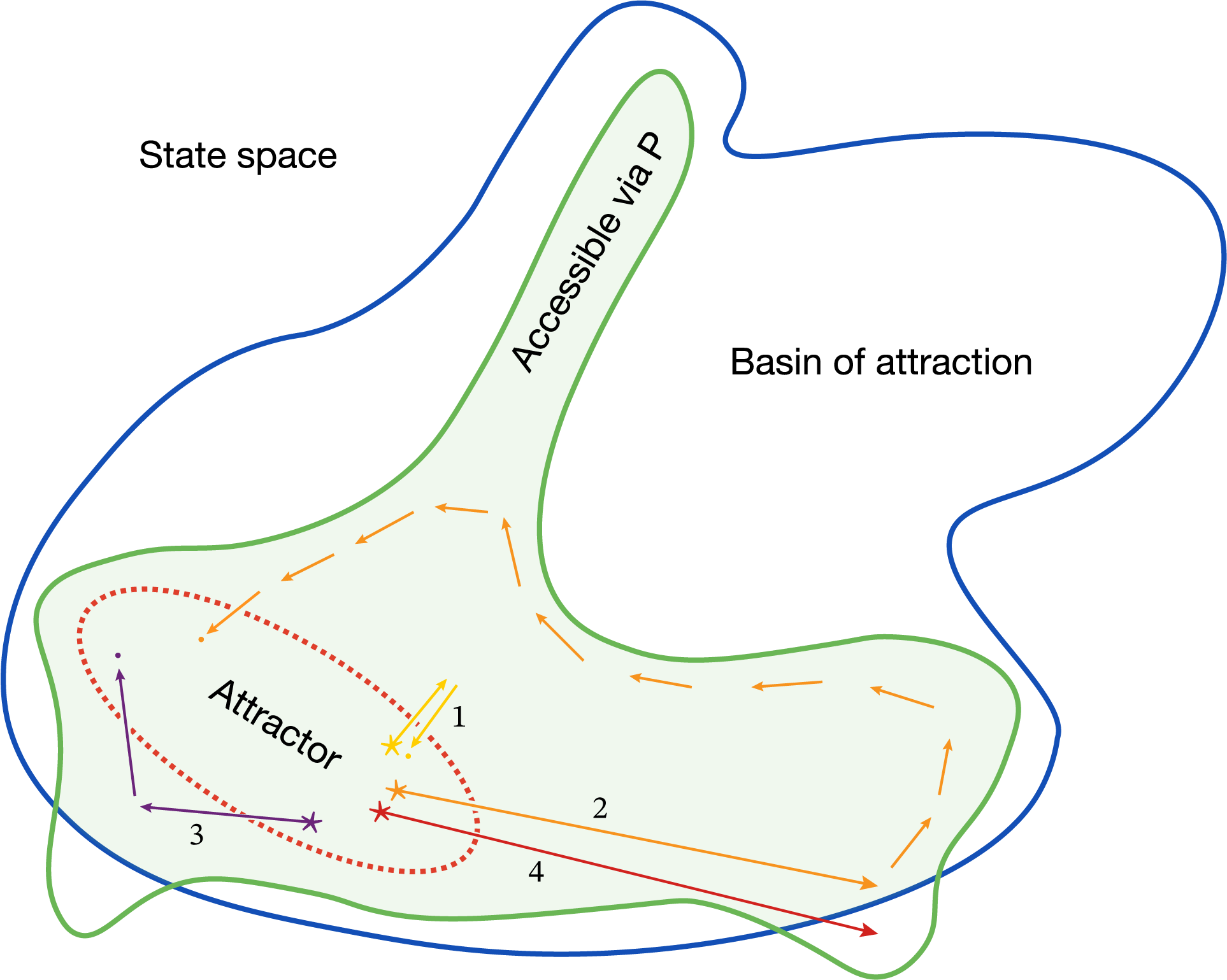}
    \caption{\textbf{Anatomy of a dynamical system under perturbation.} A schematic of state space showing an attractor (dotted red) within its basin of attraction (solid blue). The shaded green region represents states accessible via a particular perturbation $P$ -- the cognitive basin. Numbered arrows represent post-perturbation trajectories: (1) fast and unremarkable recovery; (2) slow, distorted recovery to a different region of the attractor (the free variable shifts); (3) fast recovery to a different region of the attractor; (4) exit from the basin (death). Not all states in the basin are reachable via $P$; what is accessible depends on the perturbation type.}
    \label{fig:basin}
\end{figure}

Lenia is a deterministic dynamical system (Figure~\ref{fig:basin}). Its state space is the set of all possible grid configurations, and its dynamics are a consequence of the iterative map in Equation~\ref{eq:update}. Within a given ruleset, certain initial conditions converge to self-maintaining patterns.

\textbf{Attractors.} Each of our creatures corresponds to an attractor within a particular ruleset. The attractor is not a single point but a continuous family of states: Lenia creatures persist across variations in position and heading, fluctuating slightly over time. As operationalized here, a creature's attractor is its \textit{neighborhood} (Figure~\ref{fig:basin}, dotted red). This framing makes explicit what is implicit in the Lenia literature: the search for new Lenia creatures is equivalently a search for stable attractors in the space of rulesets and grid configurations. The patterns we call creatures are not fixed objects placed on a grid but self-maintaining attractors under the dynamics induced by some update rule.

The dynamics of our creatures admit continuous symmetries -- translation and rotation -- that give the attractor the structure of a low-dimensional manifold within the full state space. These symmetries give rise to \textit{free variables}: dimensions along which the system can change while remaining in the attractor. A creature at any position, facing any direction, is the `same' creature. Because our creatures translate, heading determines their direction of travel. Heading is thus a free and consequential variable: a creature's indifference to its own heading is what affords it the potential for navigational competence in our environments (Figures~\ref{fig:env_competence} and~\ref{fig:trajectory_overlay}).

\textbf{Basin of attraction.} The basin of attraction (also called the `attractor basin') is the set of all states that converge to the attractor (Figure~\ref{fig:basin}, blue boundary). For a Lenia creature, this is the set of morphologies that, when sufficiently stepped forward through time, return to the creature's neighborhood. A full characterization of a creature's basin of attraction would reveal every perturbation it can survive. For Lenia, where the state space is continuous and high-dimensional, exhaustive enumeration is intractable.

\textbf{Cognitive domain.} Perturbations to a creature's state push it away from its attractor; a subset relax back to its canonical morphology. This is Maturana and Varela's \textit{cognitive domain}: the set of survivable perturbations \citep{Maturana1980}. In principle, all points in the basin may be obtained by perturbations in the cognitive domain.

\textbf{Cognitive basin.} Restricting to a single perturbation type, as with our targeted occlusions, accesses a subset of the basin of attraction. We call this the \textit{cognitive basin} of a perturbation $P$: one slice of the attractor basin (Figure~\ref{fig:basin}). We observe how recovery unfolds in this slice: how long it takes, how far the creature strays morphologically, and whether the consequential free variable -- heading -- shifts (Figure~\ref{fig:sweeps}).

\subsection{Targeted perturbations signal basin structure}

The targeted occlusions of the previous section map the geometry of each creature's cognitive basin. Lethal zones (red in Figure~\ref{fig:sweeps}) are perturbation locations that push the system outside the basin of attraction (trajectory 4 in Figure~\ref{fig:basin}). Quiet zones -- short recovery, small distortion, little heading change -- correspond to perturbations deep in the basin interior, where relaxation back to the attractor is fast and uneventful (trajectory 1); motion along symmetry directions remains weak, yielding only gradual reorientation. Between them lie zones of maximal reorientation: perturbations from which the creature recovers via extended, distorted paths, which produce a shift in the free variable (trajectory 2). In Lenia, we do not observe fast, undistorted recoveries that shift the free variable significantly (trajectory 3).

Zones that provoke significant reorientation lie adjacent to lethal zones on the creature's body (Figure~\ref{fig:sweeps}). This spatial proximity suggests proximity in state space: these perturbations send the system along trajectories that approach the basin boundary before relaxing back to the attractor. It is only along these near-boundary paths that we observe significant shifts in heading (Figures~\ref{fig:sweeps} and~\ref{fig:long_windy_recovery}).

Navigational capacity thus emerges near the boundary of the basin of attraction, consistent with dynamics near a separatrix in state space, where trajectories diverge sensitively between recovery, death, and other outcomes like metamorphosis and explosion. Characterizing the basin boundary directly remains an open, computationally difficult problem. More generally, these features suggest that symmetries and basin geometry jointly shape the system’s response to perturbation, with links to symmetry breaking and phase transitions (see Discussion).

\subsection{Basin geometry underlies navigational competence}

Agnosiophobia requires two properties of the creature's basin geometry. First, heading must be a free variable -- the attractor needs a dimension along which the creature can change without changing with respect to itself. Second, the creature's cognitive basin must couple perturbations to transitions along this free variable. S1s illustrates the distinction: heading is a free variable, but its basin dynamics only weakly route perturbations to reorientation (Figure~\ref{fig:sweeps}).

In our environments, creatures whose sensitivity maps grade from reorientation to lethality along the direction of travel gain both time and mechanism to redirect before occlusions become fatal. Frontal reorientation zones access critical near-boundary trajectories before the creature is sent to a different attractor. S1s again illustrates the failure case: its lethal zone, at its leading edge, leaves no buffer for reorientation. O2u's reorientation zones shield its lethal core; K4s is analogous, though some recoveries pass through an intermediate state that produces large heading change; K6s, with no lethal zones at this perturbation scale, is never pushed close enough to a lethal boundary for significant reorientation to occur (Figure~\ref{fig:trajectory_overlay}).

Our environments present obstacles that push creatures close to the basin boundary, where heading may shift. The navigational capacity we observe in Figures~\ref{fig:env_competence} and~\ref{fig:trajectory_overlay}, in this sense, follows from the basin geometry described in this section. That for our creatures this shift constitutes a competence, directing away from the occlusion, rather than toward it, is not given; it is an immutable property inherent to each creature that could have been otherwise.

\section{Discussion}

We introduced a principled perturbation environment to Lenia: regions of the grid from which no information is available. In our test creatures (Figure~\ref{fig:creatures}), we observed species-specific navigational behaviors that resemble avoidance. Creatures tend to redirect away from regions that, if encountered directly, could degrade or kill them, turning away from the perturbed flank in a manner reminiscent of damage-avoidance in amoeba. O2u (the Orbium) circumvents occluded regions, K4s collapses to an intermediate form and re-emits in the opposite direction, K6s skirts along the edges of occluded regions, and S1s dies. This avoidance arises not from an explicit representation of danger but from the same dynamics that maintain the creature's morphology. We call this behavior `agnosiophobia.’ 

To understand why and how these responses differ, we swept small, persistent occlusions across every pixel on each creature’s body. The resulting sensitivity maps (Figure~\ref{fig:sweeps}) reveal that zones sensitive to reorientation abut lethal ones, forming a gradient. Perturbations closer to lethal regions are those capable of inducing more significant changes in heading, via longer, more morphologically distorted recovery paths (Figure~\ref{fig:long_windy_recovery}). Fragility and navigational capacity are in this sense coupled.

These findings are reminiscent of spontaneous symmetry breaking and phase transitions \citep{Goldenfeld1992, Callen1974, Anderson1972}. The rules of Lenia are translationally and rotationally symmetric, yet each creature maintains a particular position and heading, constituting a symmetry-broken state. The resulting attractor is not a point but a manifold structured by these symmetries: translational symmetry is continuously expressed in the creature’s motion, while rotational symmetry defines a soft direction (one along which the system moves with little resistance), with heading as a free variable. Motion along this direction is weakly constrained: small perturbations can induce gradual reorientation, whereas perturbations that approach the basin boundary produce larger shifts. The associated increase in recovery time and distortion is consistent with critical slowing down \citep{Scheffer2009}. Reorientation reflects motion along a symmetry direction of the attractor; lethal perturbations transition to a different dynamical regime. However, we emphasize that this analogy remains qualitative: Lenia lacks an explicit thermodynamic foundation.

\subsection{Partial equifinality}

Viewing Lenia as a dynamical system, we identify two requirements for agnosiophobia. \textbf{First}, the attractor must have free variables, dimensions along which the creature may change with respect to its environment but not with respect to itself. Lenia's toroidal grid is spatially homogeneous: neither position nor heading is privileged, and may vary while morphological identity persists. \textbf{Second}, the basin dynamics must couple perturbations to transitions along these free variables. These two requirements, together with a spatial organization of sensitivity where reorientation zones precede lethal ones along the direction of travel, explain agnosiophobia.

S1s has the free variables but lacks the coupling. O2u, K4s, and K6s satisfy both, though their agnosiophobia differs qualitatively. Hence, free variables do not guarantee competence; they afford the potential for it, realized only through the right basin geometry and environment.

The mechanism underlying these competences is \textit{partial equifinality}. Equifinality is the property that many different initial conditions converge to the same final state \citep{Bertalanffy1968}. Our creatures are equifinal with respect to morphology, but the heading they recover to depends on the perturbation, and is hence non-equifinal. For our creatures, this selective equifinality enables a navigation mechanism. 

This may generalize. Free variables are continuous symmetries of the dynamics -- translation and rotation in Lenia, but in principle, any set of symmetries that a substrate admits. When perturbed, a system can `offload' stress by shifting along these free variables rather than striving to fully `reverse' a perturbation. Any dynamical system whose attractor contains free variables, and whose basin dynamics couple perturbations to transitions along those variables, has the raw material for this kind of latent competence. The more free variables an attractor contains, the more equally-valid states across which a system can distribute the cost of a perturbation. This may widen the range of novel environments in which those recoveries prove adaptive.

\subsection{Limitations}

We analyzed four non-oscillating motile creatures under one perturbation type at one spatial scale across ten environments. Future work will reveal how additional Lenia patterns handle informational occlusions and other environmental disturbances. It is not yet known why some rulesets produce coupling between perturbation and heading change while others do not. What remains is the capacity to predict, from a ruleset alone, whether a creature will exhibit agnosiophobia, as part of a roadmap for understanding the cognitive properties of novel, exotic active agents.

\subsection{Goal-directedness}

\citet{Heylighen2023} proposes that goal-directedness can be understood as convergence to an attractor: a system is goal-directed to the extent that perturbations do not push it outside its basin of attraction. He demonstrates that the classical criteria for goal-directedness -- equifinality, persistence, plasticity, and concerted action -- follow naturally from this framework. This formulation is powerful in its grounding of goal-directedness in mathematics, rather than in intuition or biological specifics. Heylighen restricts goal-directed status to far-from-equilibrium systems, motivating this class distinction by arguing that a ball returning to the bottom of a bowl does not demonstrate `what we would intuitively see as goal-directedness' \citep[p.~374]{Heylighen2023}. This provides a separation between `trivial' ball-in-bowl systems and more intuitively goal-directed systems, like human beings. 

This restriction to far-from-equilibrium systems excludes Lenia, which satisfies many of the classic criteria for goal-directedness but lacks an overt `energy' variable. More broadly, it risks excluding any system whose relationship to energy processing is unconventional, obscured, or absent, regardless of the richness of its behavioral repertoire. For our Lenia creatures, the attractor is the self-maintaining morphological pattern and thus, perhaps, the goal: to persist as a recognizable and behaviorally-consistent pattern.

It is possible that the dynamical properties of a system, some of which we describe -- attractor dimensionality, coupling to free variables, the geometry near the basin boundary, among others -- alone contain sufficient information to derive a measure of goal-directedness. Such a formalism would make room for characterizing diverse and unconventional intelligences as goal-directed and place them on a continuum. Far-from-equilibrium systems would be classified based on the complexity of their attractor landscapes rather than their thermodynamics, sharing a spectrum with systems like the ball-in-a-bowl and creatures in Lenia. This would be particularly valuable for describing multi-scale competency architectures that comprise orders of layered cognition \citep{Dennett2020, Levin2022, Kaygisiz2025, Reber2021, Lyon2006}.

\medskip\centerline{\ensuremath{\diamond}}\medskip

The agnosiophobia documented in this paper is not designed or evolved for. This adaptive and perhaps goal-directed behavior emerges from the geometry of a system’s attractor basin. Whether basin geometry alone can ground a general measure of goal-directedness remains open. For Lenia creatures, competence comes not only from the capacity to recover, but from the freedom to recover differently.

\medskip

\paragraph{Acknowledgments.} J.C. thanks Alexander Wolff Herz, Hamed Hekmat, and Gaspard Loeillot for their helpful comments and conversations. S.P. acknowledges support from the Burroughs Wellcome Fund Career Awards at the Scientific Interface. M.L. gratefully acknowledges support of Eugene Jhong and Karen Fries. We thank Marsa Hickey for the suggestion of agnosiophobia as a term for the observed behavior.

\bibliographystyle{apalike}
\bibliography{refs}

@article{Anderson1972,
  title = {More {{Is Different}}: {{Broken}} Symmetry and the Nature of the Hierarchical Structure of Science.},
  shorttitle = {More {{Is Different}}},
  author = {Anderson, P. W.},
  year = 1972,
  month = aug,
  journal = {Science},
  volume = {177},
  number = {4047},
  pages = {393--396},
  issn = {0036-8075, 1095-9203},
  doi = {10.1126/science.177.4047.393},
  urldate = {2026-04-11},
  langid = {english}
}

@article{Baluška2016,
  title = {On {{Having No Head}}: {{Cognition}} throughout {{Biological Systems}}},
  shorttitle = {On {{Having No Head}}},
  author = {Balu{\v s}ka, Franti{\v s}ek and Levin, Michael},
  year = 2016,
  month = jun,
  journal = {Frontiers in Psychology},
  volume = {7},
  publisher = {Frontiers},
  issn = {1664-1078},
  doi = {10.3389/fpsyg.2016.00902},
  urldate = {2026-04-06},
  langid = {english}
}

@article{Barandiaran2009,
  title = {Defining {{Agency}}: {{Individuality}}, {{Normativity}}, {{Asymmetry}}, and {{Spatio-temporality}} in {{Action}}},
  shorttitle = {Defining {{Agency}}},
  author = {Barandiaran, Xabier E. and Di Paolo, Ezequiel and Rohde, Marieke},
  year = 2009,
  month = oct,
  journal = {Adaptive Behavior},
  volume = {17},
  number = {5},
  pages = {367--386},
  publisher = {SAGE Publications Ltd STM},
  issn = {1059-7123},
  doi = {10.1177/1059712309343819},
  urldate = {2026-04-01},
  langid = {english}
}

@article{Beer2014,
  title = {The {{Cognitive Domain}} of a {{Glider}} in the {{Game}} of {{Life}}},
  author = {Beer, Randall D.},
  year = 2014,
  month = apr,
  journal = {Artificial Life},
  volume = {20},
  number = {2},
  pages = {183--206},
  issn = {1064-5462},
  doi = {10.1162/ARTL_a_00125},
  urldate = {2025-10-08}
}

@book{Bertalanffy1968,
  title = {General System Theory: {Foundations}, {Development}, {Applications}},
  shorttitle = {General System Theory},
  author = {von Bertalanffy, Ludwig},
  year = 1968,
  publisher = {Braziller},
  address = {New York},
  isbn = {978-0-8076-0452-6 978-0-8076-0453-3},
  langid = {english}
}

@book{Braitenberg2004,
  title = {Vehicles: {Experiments} in {Synthetic Psychology}},
  shorttitle = {Vehicles},
  author = {Braitenberg, Valentin},
  year = 2004,
  series = {Bradford Book Psychology},
  edition = {9. print},
  publisher = {MIT Press},
  address = {Cambridge, Mass.},
  isbn = {978-0-262-02208-8 978-0-262-52112-3},
  langid = {english}
}

@article{Callen1974,
  title = {Thermodynamics as a Science of Symmetry},
  author = {Callen, Herbert},
  year = 1974,
  month = dec,
  journal = {Foundations of Physics},
  volume = {4},
  number = {4},
  pages = {423--443},
  issn = {0015-9018, 1572-9516},
  doi = {10.1007/BF00708519},
  urldate = {2026-04-11},
  copyright = {http://www.springer.com/tdm},
  langid = {english}
}

@article{Chan2019,
  title = {Lenia: {{Biology}} of {{Artificial Life}}},
  shorttitle = {Lenia},
  author = {Chan, Bert Wang-Chak},
  year = 2019,
  journal = {Complex Systems},
  volume = {28},
  number = {3},
  urldate = {2025-10-01},
  langid = {american}
}

@inproceedings{Chan2020,
  title = {Lenia and {{Expanded Universe}}},
  booktitle = {{{ALIFE}} 2020: {{The}} 2020 {{Conference}} on {{Artificial Life}}},
  author = {Chan, Bert Wang-Chak},
  year = 2020,
  month = jul,
  pages = {221--229},
  publisher = {MIT Press},
  doi = {10.1162/isal_a_00297},
  urldate = {2025-10-08},
  langid = {english}
}

@article{Clawson2023,
  title = {Endless Forms Most Beautiful 2.0: {Teleonomy} and the {Bioengineering} of {Chimaeric} and {Synthetic Organisms}},
  shorttitle = {Endless Forms Most Beautiful 2.0},
  author = {Clawson, Wesley P and Levin, Michael},
  year = 2023,
  month = aug,
  journal = {Biological Journal of the Linnean Society},
  volume = {139},
  number = {4},
  pages = {457--486},
  issn = {0024-4066},
  doi = {10.1093/biolinnean/blac073},
  urldate = {2026-04-01}
}

@article{Conway1970,
  title = {Mathematical {{Games}} - {{The}} Fantastic Combinations of {{John Conway}}'s New Solitaire Game "Life"},
  author = {Conway, John},
  year = 1970,
  journal = {Scientific American},
  volume = {223},
  number = {4},
  pages = {120--123},
  publisher = {Scientific American},
  langid = {english}
}

@inproceedings{Cvjetko2025,
  title = {Discovering and {{Controlling Diverse Self-Organised Patterns}} in {{Cellular Automata Using Autotelic Reinforcement Learning}}},
  booktitle = {Alife 2025 - {{Conference}} on {{Artificial Life}}},
  author = {Cvjetko, Marko and Hamon, Gautier and {Moulin-Frier}, Cl{\'e}ment and Oudeyer, Pierre-Yves},
  year = 2025,
  month = oct,
  address = {Kyoto, Japan},
  urldate = {2026-04-03}
}

@inproceedings{Davis2024,
  title = {Non-{{Platonic Autopoiesis}} of a {{Cellular Automaton Glider}} in {{Asymptotic Lenia}}},
  booktitle = {The 2024 {{Conference}} on {{Artificial Life}}},
  author = {Davis, Q. Tyrell},
  year = 2024,
  eprint = {2407.21086},
  primaryclass = {nlin},
  doi = {10.1162/isal_a_00786},
  urldate = {2026-02-04},
  archiveprefix = {arXiv},
  langid = {english}
}

@book{Dennett1989,
  title = {The Intentional Stance},
  author = {Dennett, Daniel Clement},
  year = 1989,
  publisher = {Cambridge, Mass. : MIT Press},
  urldate = {2025-10-01},
  collaborator = {{Internet Archive}},
  isbn = {978-0-262-54053-7 978-0-262-04093-8},
  langid = {english}
}

@misc{Dennett2020,
  title = {Cognition All the Way Down: {How} to {Understand Cells}, {Tissues} and {Organisms} as {Agents} with {Agendas}},
  author = {Dennett, Daniel Clement and Levin, Michael},
  year = 2020,
  month = oct,
  journal = {aeon},
  urldate = {2026-04-01},
  howpublished = {https://aeon.co/essays/how-to-understand-cells-tissues-and-organisms-as-agents-with-agendas},
  langid = {english}
}

@inproceedings{Faldor2024,
  title = {Toward {{Artificial Open-Ended Evolution}} within {{Lenia}} Using {{Quality-Diversity}}},
  booktitle = {Artificial {{Life Conference Proceedings}} ({{ALIFE}} 2024)},
  author = {Faldor, Maxence and Cully, Antoine},
  year = 2024,
  publisher = {MIT Press},
  address = {Copenhagen, Denmark}
}

@article{Fields2022,
  title = {Competency in {{Navigating Arbitrary Spaces}} as an {{Invariant}} for {{Analyzing Cognition}} in {{Diverse Embodiments}}},
  author = {Fields, Chris and Levin, Michael},
  year = 2022,
  month = jun,
  journal = {Entropy},
  volume = {24},
  number = {6},
  pages = {819},
  address = {Basel, Switzerland},
  issn = {1099-4300},
  doi = {10.3390/e24060819},
  langid = {english},
  pmcid = {PMC9222757},
  pmid = {35741540}
}

@article{Fields2025,
  title = {Thoughts and Thinkers: {{On}} the Complementarity between Objects and Processes},
  shorttitle = {Thoughts and Thinkers},
  author = {Fields, Chris and Levin, Michael},
  year = 2025,
  month = mar,
  journal = {Physics of Life Reviews},
  volume = {52},
  pages = {256--273},
  issn = {1571-0645},
  doi = {10.1016/j.plrev.2025.01.008},
  urldate = {2025-10-28}
}

@book{Goldenfeld1992,
  title = {Lectures on {{Phase Transitions}} and the {{Renormalization Group}}},
  author = {Goldenfeld, Nigel},
  year = 1992,
  edition = {1},
  publisher = {CRC Press},
  doi = {10.1201/9780429493492},
  urldate = {2026-04-11},
  isbn = {978-0-429-49349-2},
  langid = {english}
}

@article{Hamon2025,
  title = {Discovering Sensorimotor Agency in Cellular Automata Using Diversity Search},
  author = {Hamon, Gautier and Etcheverry, Mayalen and Chan, Bert Wang-Chak and {Moulin-Frier}, Cl{\'e}ment and Oudeyer, Pierre-Yves},
  year = 2025,
  month = oct,
  journal = {Science Advances},
  volume = {11},
  number = {44},
  pages = {eadp0834},
  publisher = {American Association for the Advancement of Science},
  doi = {10.1126/sciadv.adp0834},
  urldate = {2025-11-18}
}

@article{Heylighen2023,
  title = {The Meaning and Origin of Goal-Directedness: A Dynamical Systems Perspective},
  shorttitle = {The Meaning and Origin of Goal-Directedness},
  author = {Heylighen, Francis},
  year = 2023,
  month = aug,
  journal = {Biological Journal of the Linnean Society},
  volume = {139},
  number = {4},
  pages = {370--387},
  issn = {0024-4066},
  doi = {10.1093/biolinnean/blac060},
  urldate = {2025-10-05}
}

@misc{Hudcová2026,
  title = {Visualizing the {{Structure}} of {{Lenia Parameter Space}}},
  author = {Hudcov{\'a}, Barbora and Du{\v s}ek, Franti{\v s}ek and Tuccio, Marco and Hongler, Cl{\'e}ment},
  year = 2026,
  howpublished = {arXiv preprint arXiv:2601.01932},
  doi = {10.48550/arXiv.2601.01932}
}

@inproceedings{Kawaguchi2021,
  title = {Introducing Asymptotics to the State-Updating Rule in {{Lenia}}},
  booktitle = {{{ALIFE}} 2021: {{The}} 2021 {{Conference}} on {{Artificial Life}}},
  author = {Kawaguchi, Takako and Suzuki, Reiji and Arita, Takaya and Chan, Bert},
  year = 2021,
  month = jul,
  publisher = {MIT Press},
  doi = {10.1162/isal_a_00425},
  urldate = {2026-04-04},
  langid = {english}
}

@article{Kaygisiz2025,
  title = {Can {{Molecular Systems Learn}}?},
  author = {Kaygisiz, K{\"u}bra and Ulijn, Rein V.},
  year = 2025,
  journal = {ChemSystemsChem},
  volume = {7},
  number = {2},
  pages = {e202400075},
  issn = {2570-4206},
  doi = {10.1002/syst.202400075},
  urldate = {2026-04-05},
  copyright = {\copyright{} 2025 Wiley-VCH GmbH},
  langid = {english}
}

@article{Kriegman2020,
  title = {A Scalable Pipeline for Designing Reconfigurable Organisms},
  author = {Kriegman, Sam and Blackiston, Douglas and Levin, Michael and Bongard, Josh},
  year = 2020,
  month = jan,
  journal = {Proceedings of the National Academy of Sciences of the United States of America},
  volume = {117},
  number = {4},
  pages = {1853--1859},
  issn = {0027-8424},
  doi = {10.1073/pnas.1910837117},
  urldate = {2026-04-06},
  pmcid = {PMC6994979},
  pmid = {31932426}
}

@article{Kumar2025,
  title = {Automating the {{Search}} for {{Artificial Life}} with {{Foundation Models}}},
  author = {Kumar, Akarsh and Lu, Chris and Kirsch, Louis and Tang, Yujin and Stanley, Kenneth O. and Isola, Phillip and Ha, David},
  year = 2025,
  journal = {Artificial Life},
  volume = {31},
  number = {3},
  pages = {368--396}
}

@book{Küppers1990,
  title = {Information and the {{Origin}} of {{Life}}},
  author = {K{\"u}ppers, Bernd-Olaf},
  year = 1990,
  month = apr,
  publisher = {MIT Press},
  address = {Cambridge, MA, USA},
  isbn = {978-0-262-11142-3},
  langid = {english}
}

@article{Levin2019,
  title = {The {{Computational Boundary}} of a ``{{Self}}'': {{Developmental Bioelectricity Drives Multicellularity}} and {{Scale-Free Cognition}}},
  shorttitle = {The {{Computational Boundary}} of a ``{{Self}}''},
  author = {Levin, Michael},
  year = 2019,
  month = dec,
  journal = {Frontiers in Psychology},
  volume = {10},
  publisher = {Frontiers},
  issn = {1664-1078},
  doi = {10.3389/fpsyg.2019.02688},
  urldate = {2025-10-01},
  langid = {english}
}

@article{Levin2022,
  title = {Technological {{Approach}} to {{Mind Everywhere}}: {{An Experimentally-Grounded Framework}} for {{Understanding Diverse Bodies}} and {{Minds}}},
  shorttitle = {Technological {{Approach}} to {{Mind Everywhere}}},
  author = {Levin, Michael},
  year = 2022,
  month = mar,
  journal = {Frontiers in Systems Neuroscience},
  volume = {16},
  publisher = {Frontiers},
  issn = {1662-5137},
  doi = {10.3389/fnsys.2022.768201},
  urldate = {2025-10-01},
  langid = {english}
}

@article{Lyon2006,
  title = {The Biogenic Approach to Cognition},
  author = {Lyon, Pamela},
  year = 2006,
  month = mar,
  journal = {Cognitive Processing},
  volume = {7},
  number = {1},
  pages = {11--29},
  issn = {1612-4782},
  doi = {10.1007/s10339-005-0016-8},
  langid = {english},
  pmid = {16628463}
}

@book{Maturana1980,
  title = {Autopoiesis and {{Cognition}}: {{The Realization}} of the {{Living}}},
  shorttitle = {Autopoiesis and {{Cognition}}},
  author = {Maturana, Humberto R. and Varela, Francisco J.},
  year = 1980,
  series = {Boston {{Studies}} in the {{Philosophy}} and {{History}} of {{Science}}},
  volume = {42},
  publisher = {Springer Netherlands},
  address = {Dordrecht},
  doi = {10.1007/978-94-009-8947-4},
  urldate = {2026-04-03},
  copyright = {http://www.springer.com/tdm},
  isbn = {978-90-277-1016-1 978-94-009-8947-4},
  langid = {english}
}

@misc{Mirmomeni2014,
  title = {Is Information a Selectable Trait?},
  author = {Mirmomeni, Masoud and Punch, William F. and Adami, Christoph},
  year = 2014,
  howpublished = {arXiv preprint arXiv:1408.3651},
  doi = {10.48550/arXiv.1408.3651}
}

@article{Reber2021,
  title = {Cognition in Some Surprising Places},
  author = {Reber, Arthur S. and Balu{\v s}ka, Franti{\v s}ek},
  year = 2021,
  month = jul,
  journal = {Biochemical and Biophysical Research Communications},
  volume = {564},
  pages = {150--157},
  issn = {1090-2104},
  doi = {10.1016/j.bbrc.2020.08.115},
  langid = {english},
  pmid = {32950231}
}

@inproceedings{Reinke2020,
  title = {Intrinsically {{Motivated Discovery}} of {{Diverse Patterns}} in {{Self-Organizing Systems}}},
  booktitle = {International {{Conference}} on {{Learning Representations}} ({{ICLR}})},
  author = {Reinke, Chris and Etcheverry, Mayalen and Oudeyer, Pierre-Yves},
  year = 2020,
  address = {Addis Ababa, Ethiopia}
}

@article{Rosenblueth1943,
  title = {Behavior, {{Purpose}} and {{Teleology}}},
  author = {Rosenblueth, Arturo and Wiener, Norbert and Bigelow, Julian},
  year = 1943,
  journal = {Philosophy of Science},
  volume = {10},
  number = {1},
  pages = {18--24},
  publisher = {[Cambridge University Press, The University of Chicago Press, Philosophy of Science Association]},
  issn = {0031-8248},
  urldate = {2026-04-01}
}

@article{Scheffer2009,
  title = {Early-Warning Signals for Critical Transitions},
  author = {Scheffer, Marten and Bascompte, Jordi and Brock, William A. and Brovkin, Victor and Carpenter, Stephen R. and Dakos, Vasilis and Held, Hermann and {van Nes}, Egbert H. and Rietkerk, Max and Sugihara, George},
  year = 2009,
  month = sep,
  journal = {Nature},
  volume = {461},
  number = {7260},
  pages = {53--59},
  publisher = {Nature Publishing Group},
  issn = {1476-4687},
  doi = {10.1038/nature08227},
  urldate = {2026-04-11},
  copyright = {2009 Macmillan Publishers Limited. All rights reserved},
  langid = {english}
}

@book{Shanahan2010,
  title = {Embodiment and the Inner Life: {{Cognition}} and {{Consciousness}} in the {{Space}} of {{Possible Minds}}},
  shorttitle = {Embodiment and the Inner Life},
  author = {Shanahan, Murray},
  year = 2010,
  month = jun,
  publisher = {Oxford University Press},
  doi = {10.1093/acprof:oso/9780199226559.001.0001},
  urldate = {2026-04-06},
  isbn = {978-0-19-922655-9}
}

@incollection{Sloman1984,
  title = {The {{Structure}} of the {{Space}} of {{Possible Minds}}},
  booktitle = {The {{Mind}} and the {{Machine}}: Philosophical Aspects of {{Artificial Intelligence}}},
  author = {Sloman, Aaron},
  editor = {Torrance, Stephen B.},
  year = 1984,
  publisher = {Ellis Horwood},
  address = {Chichester},
  urldate = {2026-04-01},
  langid = {english}
}

@article{Srivastava2014,
  title = {Dropout: {{A Simple Way}} to {{Prevent Neural Networks}} from {{Overfitting}}},
  shorttitle = {Dropout},
  author = {Srivastava, Nitish and Hinton, Geoffrey and Krizhevsky, Alex and Sutskever, Ilya and Salakhutdinov, Ruslan},
  year = 2014,
  journal = {Journal of Machine Learning Research},
  volume = {15},
  number = {56},
  pages = {1929--1958},
  issn = {1533-7928},
  urldate = {2026-04-04}
}

@article{Zhang2025,
  title = {Classical Sorting Algorithms as a Model of Morphogenesis: {{Self-sorting}} Arrays Reveal Unexpected Competencies in a Minimal Model of Basal Intelligence},
  shorttitle = {Classical Sorting Algorithms as a Model of Morphogenesis},
  author = {Zhang, Taining and Goldstein, Adam and Levin, Michael},
  year = 2025,
  month = feb,
  journal = {Adaptive Behavior},
  volume = {33},
  number = {1},
  pages = {25--54},
  publisher = {SAGE Publications Ltd STM},
  issn = {1059-7123},
  doi = {10.1177/10597123241269740},
  urldate = {2025-12-11},
  langid = {english}
}

\end{document}